\documentclass[authoryear,5p]{elsarticle}
\usepackage{graphicx}
\usepackage{paralist}
\usepackage{mathptmx}      % use Times fonts if available on your TeX system
\usepackage{hyperref}
\usepackage[svgnames]{xcolor}  %http://mirrors.ctan.org/macros/latex/contrib/xcolor/xcolor.pdf
\usepackage{todonotes}

\usepackage[utf8]{inputenc} % Allow Umlauts in input
\usepackage[T1]{fontenc} % Use T1 encoding so that copy and paste works
\usepackage[autolanguage]{numprint} % printing numbers with thousand-separators
\usepackage{dirtytalk} % nice package for quiting inteviews
\npthousandsep{\,}

\newcommand{\Rplot}[3]{
\begin{figure}[t]
\centerline{\includegraphics[scale=#3]{#1}}%
\vspace{-5mm}\caption{#2}\label{#1}%
\end{figure}}

\newcommand{\Rplotw}[3]{
\begin{figure*}[t]
\centerline{\includegraphics[scale=#3]{#1}}%
\vspace{-5mm}\caption{#2}\label{#1}%
\end{figure*}}

\newcommand{\wesay}[1]{{\color{DarkGreen}\textit{``#1''}}}
\newcommand{\theysay}[1]{{\color{Maroon}\textit{``#1''}}}
\journal{Information and Software Technology}
\begin{document}

\begin{frontmatter}

\title{A Community's Perspective on the Status and Future of\\
       Peer Review in Software Engineering}

\author[fub]{Lutz Prechelt}
\ead{prechelt@inf.fu-berlin.de}
\author[ustutt]{Daniel Graziotin}
\ead{daniel.graziotin@informatik.uni-stuttgart.de}
\author[tum]{Daniel M\'{e}ndez Fern\'{a}ndez}
\ead{daniel.mendez@tum.de}
\address[fub]{Freie Universit\"at Berlin, Berlin, Germany}
\address[ustutt]{University of Stuttgart, Stuttgart, Germany}
\address[tum]{Technical University of Munich, Garching, Germany}

\begin{abstract}
\emph{Context:}
Pre-publication peer review of scientific articles is considered 
a key element of the research process in software engineering, yet it is often perceived as not
to work fully well.
\emph{Objective:}
We aim at understanding the perceptions of and attitudes towards peer review
of authors and reviewers at one of software engineering's most prestigious venues, the
International Conference on Software Engineering (ICSE).
\emph{Method:}
We invited 932 ICSE 2014/15/16 authors and reviewers to participate in
a survey with 10 closed and 9 open questions.
\emph{Results:}
We present a multitude of results, such as:
Respondents perceive only one third of all reviews to be good,
yet one third as useless or misleading; 
they propose double-blind or zero-blind reviewing regimes for improvement; 
they would like to see showable proofs of (good) reviewing work be
introduced;
attitude change trends are weak.
\emph{Conclusion:}
The perception of the current state of software engineering peer review is
fairly negative.
Also, we found hardly any trend that suggests reviewing will improve
by itself over time; the community will have to make explicit efforts.
Fortunately, our (mostly senior) respondents appear more open for
trying different peer reviewing regimes than we had expected.
\end{abstract}

\begin{keyword}
survey, peer review, open review, double-blind review, reward
\end{keyword}

\end{frontmatter}

\tableofcontents

%========================================================================
\section{Introduction}
\label{introduction}

For our purposes, peer review is the practice by which a publication venue
sends an article to several expert colleagues (the peers) for review
before it is accepted for publication (or not).
Although a few venues recently started trying out a different approach
(e.g., \cite{F1000,ScienceOpen}), this basic model of \emph{pre-publication peer
review} is usually considered a cornerstone of quality assurance in the
scientific process, in software engineering and beyond~\citep{mulligan2013}.

This article attempts to understand what is currently working well or
not-so-well about peer review in software engineering (SE) and how this
might change in the next 20 years.

%------------------------------------------------------------------------
\subsection{Variants of Peer Review}

The acceptance decision may be made after just one round of 
reviewing (single-stage peer review\footnote{authors may be allowed to
  comment on the reviews: rebuttals}), 
typical for conferences, 
or after multiple rounds with improvements of the work
(multi-stage review\footnote{always with rebuttals}), 
typical for journals.

Usually, the authors do not know the identity of the reviewers (blind
review).
Reviewers might know the identity of the authors (single-blind review)
or not (double-blind review).
Only rarely do the authors get to know the names of reviewers
(non-blind review, zero-blind review) or does the public get to see
the content of the reviews (open review, public review).

%------------------------------------------------------------------------
\subsection{Issues with Peer Review}

Informally, researchers often criticize peer review as not doing its
job properly and indeed the practice has various inherent problems,
for instance:
\begin{itemize}
\item Reviewers will not always be competent to properly review a
  particular work, and often provide inconsistent reports
  \citep{Baxt1998,Smith2006}.
\item Reviewers will sometimes be biased against certain aspects of
  the work: methods, technology, goals, etc. \citep{Armstrong1997}.
\item Reviewers may, protected by their anonymity, abuse their power
  to inhibit the publication of lines of work that compete with their own
  \citep{Smith2006,Hettyey2012}.
\item Reviewing can be viewed as contributing little to the reviewer's
  reputation and so reviewer motivation can be lacking and reviewing
  be done rather sloppily \citep{zaharie2016}.
\end{itemize}

Because of issues like these, other fields (most prominently in the
biomedical realm) have long worked to understand the status of peer
reviewing and how to improve it \citep{Nature06}. 
For instance, such research has produced strong evidence that
double-blind reviewing will lead to results that are less biased than
with single-blind reviewing, e.g. \citet{BudTreAar08}, a fact that
is now also being picked up in software engineering \citep{Bacchelli2017}.
But beyond that, software engineering venues are not, so far,
particularly prone to experimentation with possible improvements to the peer
reviewing regime.
In light of the above issues, this might be a pity.

For instance, the high-class health journal The BMJ (acceptance rate 7\%)
not only performs reviews zero-blind (that is, reviewers sign their
reviews), they also publish the reviews along with accepted articles
(open reviewing, \citet{BMJopen}); 
there is no comparable software engineering venue
doing anything as radical.

%------------------------------------------------------------------------
\subsection{Research questions}
\label{researchquestions}
Our perspective is understanding and then improving the peer review process.
We designed our survey along the following research questions.
Results and discussion will be structured mostly into one section per
research question.

\bgroup
\newcommand{\Sec}[1]{\emph{Section #1:}}
\noindent
\Sec{5} What do authors and reviewers perceive to be the purposes of
  peer review? 
  Which are more important than which others?\\
\Sec{6} How well do they perceive peer review to work today 
  (in the sense of producing
  valid and helpful reviews) and why?\\
\Sec{7} How much should reviewers and authors be blinded?\\
\Sec{8} Which aspects of reviewing should be public?\\
\Sec{9} Should reviewers be compensated for their work? How?\\
\Sec{10} What changes to the current reviewing regime should be performed?\\
\Sec{11} How might the answers to each of the above questions change in the 
  next few decades?\\
\egroup

%------------------------------------------------------------------------
\subsection{Research contribution}
\label{researchcontribution}

Our article makes two research contributions:
First, it characterizes the attitudes of mostly senior members of the
ICSE\footnote{International Conference on Software Engineering}%, a top Software Engineering venue.} 
authors-and-reviewers community with respect to the research questions.
Second, it predicts how these attitudes will likely be different for a
similar sample of people in the future, several decades away.

%------------------------------------------------------------------------
\subsection{Structure of this article}
\label{structurearticle}

After reviewing related work (Section~\ref{relatedwork}),
we will present our method: 
The survey population (Section~\ref{population}),
the survey instrument (\ref{instrument}),
the execution of the survey (\ref{execution}),
our data analysis techniques (\ref{quant-techniques} and
\ref{qual-techniques}), and
the resulting public data archive (\ref{sec:online-material}).
Then, we discuss the respondent demographics (\ref{demographics})
before presenting the results structured according to our list of
research questions (Sections \ref{resultsfirst} to \ref{resultslast}).
We then discuss our study's limitations (Section~\ref{limitations})
before we conclude (Section~\ref{conclusions}).

%========================================================================
\section{Related work}
\label{relatedwork}

We organize this section along the research questions from 
Section \ref{researchquestions}.
What sets our study apart from other survey work in the area is the
use of open questions and qualitative analysis.
While we reference various related work, we consider two large scale
surveys of peer reviewers attitudes across many disciplines 
as our baseline background material upon which
we frame our study primarily: 
First \citet{mulligan2013} with 4037 respondents,
second \citet{RosDepSch17} with 3062.
The latter, organized by OpenAIRE, an Open Access collaboration project,
is special in that 76\% of respondents reported to have participated in
open reviewing previously; an unusual population.
We found only one reviewing study in the software engineering literature
\citep{Bacchelli2017}, also a survey.

\textbf{Purpose of peer review:}
\citet[p.xii]{weller2001editorial} proposed a concise characterization:
``The valid article is accepted, the messy article cleaned up, and the invalid article rejected''. 
The \citet{mulligan2013} survey found the main perceived purposes to be
(in this order):
to improve the quality of published papers; to determine their originality; 
to select the best possible manuscripts for a journal.
Our work will ask the question also beyond predefined answer categories
and ask for elaboration.

\textbf{How well does peer review work today:}
The \citet{mulligan2013} survey had 69\% of respondents report high or
very high satisfaction.
When asked what aspects of their articles were improved the most
through peer review, respondents mentioned the introduction most (90\%)
and statistical methods least.
Our work will ask about percentages of good, mediocre, or bad reviews
received and about specific positive and negative peer review
experiences to provide a more detailed picture.

\textbf{Blinding:}
Much discussion has happened lately on how much anonymity should be in
the peer review process~\citep{Pontille2014, jubb2016}.
Empirical research has found interesting effects from double-blind reviewing.
For instance,
\citet{BudTreAar08} found that more articles of female researchers were
  accepted after the journal Behavioral Ecology adopted
  double-blind review (but not in other journals that did not).
\citet{LabPie94} found for a sample of economics journals (and
  controlling for several confounding factors) that 
  articles accepted after single-blind review were cited less often than 
  articles accepted after double-blind review.
As for software engineering,
\citet{Bacchelli2017} survey how double-blind peer review is
perceived by the ICSE community and find that about
half of the respondents believe all software engineering venues should 
switch to double-blind reviewing.
\citet{SeeBac17} investigate bibliographic data from 71 of the 80
largest computer science conferences of 2014 and 2015 and find evidence
that newcomers (people who have not previously published at that conference)
get a smaller share of a conference when single-blind reviewing is
used compared to conferences using double-blind reviewing.

The \citet{mulligan2013} survey respondents did not like the prospect
that their names be made visible to the authors (8\% more likely to be
willing to review under such circumstances, 51\% less likely) 
or to the readers (18\% and 45\%). 
In the OpenAIRE survey, 67\% of respondents believed zero-blind reviewing would 
make reviewers less inclinced to provide a review and 
44\% believed it would improve review quality; 
65\% believed it makes strong criticism less likely \citep{RosDepSch17}.
Our study will ask for degrees of agreement with double-blinding and
zero-blinding.

\textbf{Publicness:}
Support for the review reports to be published alongside the accepted
paper was low (11\% more likely and 58\% less likely) in
\citet{mulligan2013}. 
Similar percentages were found for the possibility of disclosing names to authors only (8\% and 51\%)
and for having the reviewer names only published alongside the article (18\% and 45\%).
Even in the OpenAIRE survey,
52\% of respondents expect reviewers to become less inclined to review, although
65\%  expect published reviews to be useful for readers,
60\% expect an increase in review quality, and
45\% expect authors to become more inclined to submit to such journals.

Some venues such as 
F1000Research~\citep{F1000},
ScienceOpen~\citep{ScienceOpen}, or
The BMJ~\citep{BMJopen}
require public reviews,
and initiatives such as Publons~\citep{Publons}
or Academic Karma\footnote{\url{http://academickarma.org}}
promote them for the rest of the scientific publishing world.
Our study will ask for degrees of agreement with publicness of reviews.

\textbf{Reviewer compensation:}
%Should reviewers be compensated for their work? How?
Overall scientific publication rates are increasing by  8-9\% each
year \citep{bornmann2015}.
As a result, there is a \textit{reviewer fatigue syndrome} \citep{breuning2015}:
reviewers decline review invitations more and more often \citep{mulligan2013}.
\citet{warne2016}, a study specifically on reviewer compensation,
reports mean agreement of 4.0 (on a 1-to-5 scale,
based on 3000 surveyed researchers) with the statement ``I would spend
more time reviewing if it was recognised as a measurable research
activity''.
51\% of the \citet{mulligan2013} participants would more likely review for a
venue that compensated them somehow,
only 15\% less likely.
Our study asks for degrees of agreement and for specific compensation ideas.

\textbf{Useful reviewing regime changes:}
%What changes to the current reviewing regime should be performed?
About 30\% of the \citet{mulligan2013} respondents believed that the
current status of peer review is the best we can have, 
but the study did not ask the other 70\% for improvement suggestions.
Several such suggestions come from viewpoint articles.
For example, 
\citet{ralph2016} recommends for Information Systems research to
provide editorial review only for empirical articles and to
desk-reject many of those based on checklists.
\citet{ferreira2016} recommends to demand a rate of reviews for each
scientist, standardize peer review through training in academic
curricula and workshops, and decoupling peer review from journals.
Our study asks for any improvement idea, plus elaboration.

\textbf{Future change:}
We are not aware of any study that goes beyond reporting current attitudes
to explicitly extrapolating them into the future in a data-based manner.
Our study will do so based on regression modeling with demographic variables.

We will refer to specific similar or contrasting results of related
work as appropriate when we discuss our results.

%========================================================================
\section{Methods}
\label{methods}

Our results are based on a mixed quantitative/qualitative survey of
software engineering authors and reviewers.

%------------------------------------------------------------------------
\subsection{Survey population}
\label{population}

As our base population, we pick the set of all authors and reviewers
of recent instances of the
\emph{International Conference on Software Engineering} (ICSE 2014,
2015, and 2016), because it represents software engineering
research broadly across most topic ranges and at a high level of quality.
We collected the author email addresses from the published articles
and the reviewer addresses via the program committee web pages or from
lists provided by the program committee chairs.
Reviewers include the members of program committee (each year), 
review committee (2015 only), and program board (2014 and 2016 only).

This results in a set of 966 people.
Of these, 642 (66\%) have been an author in only one year, 
99 (10\%) were authors in multiple years. Further,
156 (16\%) served as reviewer in one year, and
68 (7\%) served as reviewer in multiple years.
Of 34 people (3.5\%), we could not produce an individual email address
(e.g., because no author address was given at all or all authors shared one address),
resulting in an actual base population of 932 people.

%------------------------------------------------------------------------
\subsection{Survey instrument and execution}
\label{instrument}
\label{execution}

Our questionnaire was built from scratch and had 19 questions.
They were a mix of closed or quantitative ones on the one hand and 
open ones for qualitative analysis on the other.
Most closed questions used a 10-point disagree/agree scale,
the others are numeric or binary.
We will often provide specific wording from the
questionnaire along with the presentation of the results.
The questionnaire is openly available (see Section \ref{sec:online-material})
We sent out an invitation email to the base population in August 2016,
stating
``We kindly ask you to participate in a small survey on the future of
peer review. 
Your participation, by answering 19 questions that take about 15
minutes of your time, will broaden the understanding of peer reviewing
specifically in software engineering:
(1) How are current reviewing practices perceived?
(2) How could the peer review process be improved?''.

The invitation contained one link to the questionnaire and another by which
a recipient could tell us s/he had left software engineering research
and would not reply for that reason.

We left the survey open for 14 days (this was mentioned in the email)
and sent no reminder.
We received 74 bounce messages from email addresses that had meanwhile 
become invalid; these will mostly belong to junior authors.
We received 45 out-of-office autoreplies, 13 of which pointed to an
absence of the recipient of more than one week.
These 74+13 cases reduce our effective base population to 845.

The ``no longer a researcher link'' was used by 19 people,
reducing it further to 826.

The survey had 241 respondents, giving a 29\% response rate.
167 respondents (69\% of 241) worked through all pages of the questionnaire.
As all questions were optional, each single question has a lower (and varying) 
number of responses. 
The time between opening the survey and answering the last question ranged
from 4 minutes to 1 day, 22 hours;
% 4 minutes, 17 seconds
the first and third quartile were 12 minutes and 
% 11 minutes, 40 seconds
% 26 minutes, 38 seconds
27 minutes, respectively.
The median completion time was 17 minutes.
% 17 minutes, 28 seconds
% mean 40m 34s

%------------------------------------------------------------------------
\subsection{Quantitative data analysis methods}
\label{quant-techniques}

For the quantitative data, we mostly report percentages relative
to the respective number of responses and
sometimes visualize it with box plots and Likert plots.
We use linear modeling for the extrapolation into the future.

%------------------------------------------------------------------------
\subsection{Qualitative data analysis methods}
\label{qual-techniques}

For the open questions, we applied a rough version of open coding
\citet[II.5]{StrCor90} to derive a reasonable post-hoc classification
of the responses so that we can quantify the frequency of the most
common types of response.
We kept the coding process simple: There was no pre-specified
granularity goal or semantic styles goal for the codes, nor did we
align codes across different questions.
Many codes occurred only rarely, we will therefore not present, define, or
even mention all codes.

%------------------------------------------------------------------------
\subsection{Data-and-materials archive}
\label{sec:online-material}
To increase the transparency of our study and its reproducibility, 
we disclose all the instruments used and the data analyzed in an online open science archive including:
\begin{compactitem}
\item a README file,
\item the questionnaire, 
\item the raw collected data (including the answers to the open questions), 
\item the results of the open coding (annotations and codebooks),
\item the statistical and plotting routines, 
\item and the resulting plots. 
\end{compactitem}
The archive can be found in the open science repository of the present paper
~\citep{Prechelt2017}.

%========================================================================
\section{Results: Demographics}
\label{demographics}
\label{resultsfirst}

Compared to the population, our respondents ($n=141$ for this
question, which is treated as 100\% for this question) are extremely senior.
52\% identified themselves as tenured professors,
15\% as non-tenured professors, and
22\% as being on the post-doctoral level or ``industrial researcher'' level.

Of those who provided a gender ($n=143$, 100\%),
15\% identified themselves as female,
85\% as male.

Of those who provided an age ($n=140$, 100\%),
14\% were in their twenties (minimum: 23),
43\% in their thirties,
21\% in their fourties,
17\% in their fifties, and
4\% beyond (maximum: 67).

Those who stated their country of affiliation ($n=157$, 100\%),
come from 32 different countries, the most common being
USA (33\%), Germany (12\%), and Canada (6\%), and all others below 5\%.

Respondents said they had published 6.2 peer-reviewed articles 
in the past twelve months
and 1.6 articles at the three ICSEs in question, on average.
They had been reviewers at 0.8 of those three ICSEs on average.
For the base population, the latter value is 0.3; another indication of our
respondents' strong seniority.

In the results below, we will report on different
subgroups where appropriate.

%========================================================================
\section{Results: Purpose of peer review }\label{purpose}
%columns C to V

We asked respondents how much they agree with each of the nine
suggested purposes of peer review shown in Figure~\ref{purpose.pdf}.

\Rplotw{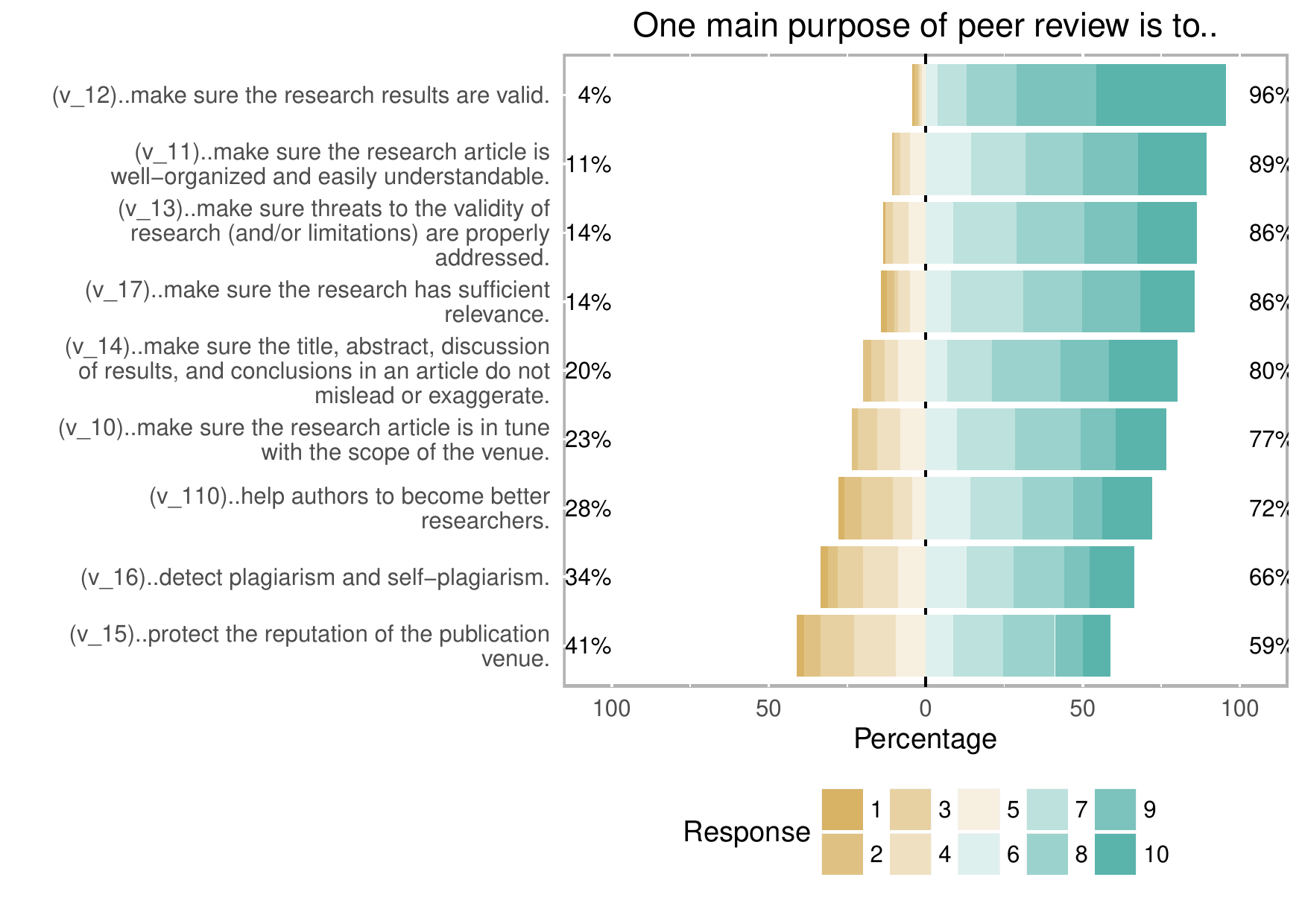}{Purposes of peer review.
  Answers range from strong disagreement (1) to strong agreement (10).}{0.9}

All nine purposes receive more than 50\% of replies on the
``agree''-side of the scale, six of them even more than 75\%.
The, by far, most popular answer is to ensure the validity of the research, the
core of peer review's gatekeeping function.
The runners-up are to make sure the article is well written,
limitations are properly discussed, and the research is relevant.
Relatively least popular are detecting plagiarism and protecting the
reputation of the venue.

The question was followed by three open text slots to add additional
concerns in the form of open answers.

Open coding of the open-ended answers found 13 categories.
The top two (each occurring in 17 of the responses) refer to
ensuring the novelty of the results and to ensuring scientific progress,
respectively.
Some of the novelty-related answers stressed specific aspects,
such as \theysay{[\ldots] not just in
  the ICSE community but in the broader research
  community}\footnote{Our response quotations are usually verbatim excerpts,
  except for spelling and punctuation corrections.
  If we applied changes to wording (for comprehensibility or anonymization),
  these are indicated by square brackets.}
or \theysay{Ensure that innovative, but possibly incomplete, ideas are
  injected into the community to stimulate discovery and innovation.}
Ensuring progress was characterized in many different ways, from general ones 
(\theysay{assessing contribution to the field})
down to rather specific aspects such as \theysay{To ensure that the reporting
allows for reproducibility and replicability}.

The third-most frequent code (occurring 13 times) represents
checking that 
articles make proper use of related work, 
relate themselves to the state of the art, and 
provide appropriate theoretical framing of their research design.

Most of the other codes (occurring between 10 times and 2 times) echo
concerns already represented in the categories of 
Figure \ref{purpose.pdf}, but the respondents added detail or
emphasized a sub-aspect.
Most popular among those were ensuring \theysay{quality} (10 occurrences)
or \theysay{soundness} (e.g. of method execution or result interpretation,
9 occurrences),
improving writing (10 occurrences), and
\theysay{selecting} among articles (e.g. \theysay{grain from chaff}, 
\theysay{top contributions}, or \theysay{To balance acceptances across topic areas}, 
9 occurrences).

Two of the other codes, however, are new:
Learning (from other reviewers or about current research, 5
occurrences) and
ensuring impact on SE practice (2 occurrences).

\cite{mulligan2013} found that the purpose of peer review is, respectively,
to improve the quality of published papers (94\%); 
to determine their originality (92\%);
and to select the best possible manuscripts for a journal (85.5\%).
We are not aware of studies exploring the purpose of peer review in an open-ended way,
as we did.
Our results offer a base for future studies on the purpose of peer review.

%========================================================================
\section{Results: How well does peer-review work?}
\label{quality}

%------------------------------------------------------------------------
\subsection{Quantitative estimate}
\label{how-well-quantitative}
%columns W-Z

It is problematic to ask for the net effect of peer reviewing unless
the population consists exclusively of editors or PC chairs.
So instead we asked \wesay{As an author, what percentage of the reviews
  that you receive is 
  good, reasonable, unhelpful or grossly faulty.} 
and elaborated as follows:
\begin{itemize}
\item \wesay{By 'good' we mean a review that is helpful for the acceptance
  decision and for the authors and that substantiates all of its
  important points of criticism or praise. It may be quite critical
  and even propose rejection.}
\item \wesay{By 'reasonable' we mean a review that is 'good' in some
  respects, but lacks detail in others.}
\item \wesay{By 'unhelpful' we mean a review that largely or completely lacks
  substance.}
\item \wesay{By 'grossly faulty' we mean a review that misunderstands or
  ignores key aspects of the article, leading to wildly exaggerated
  praise or criticism; this covers only questions of fact, not of
  opinion or weighting.}
\end{itemize}

\Rplot{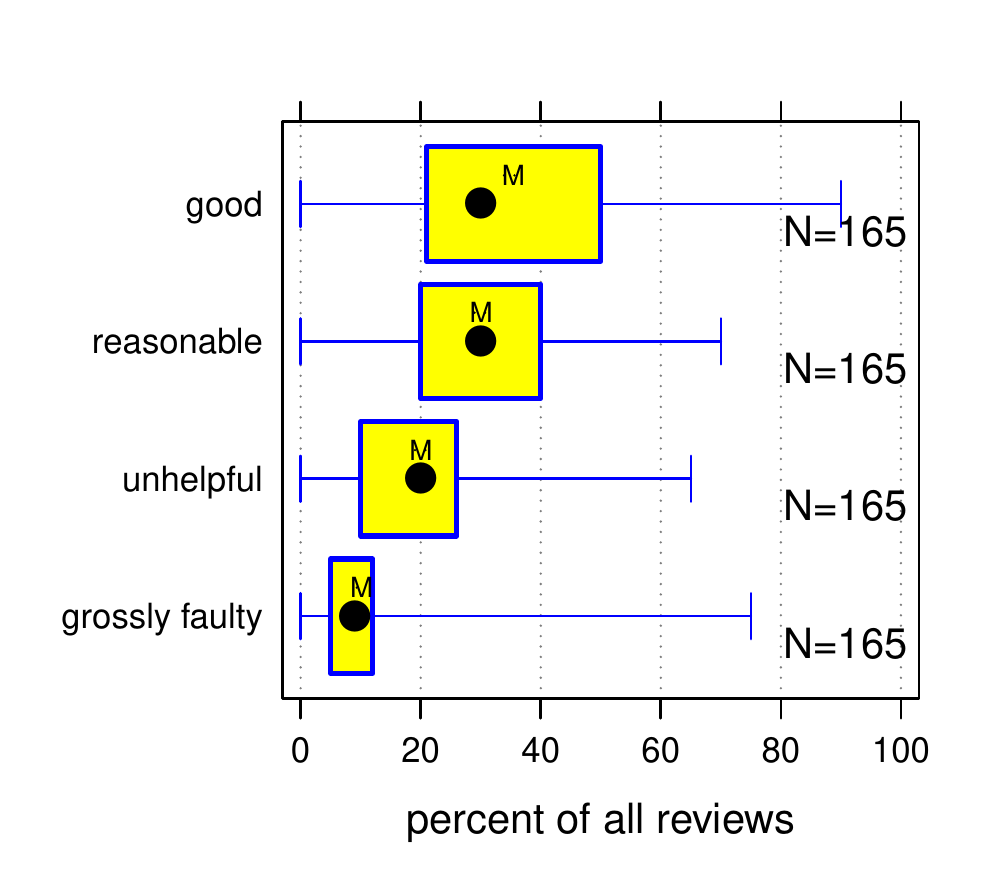}{Answer distributions for the frequency of
  four quality levels of peer reviews.
  Whiskers show minimum/maximum, the fat dot is the median, the M the
  arithmetic mean.}{0.9} 

The results are shown in Figure~\ref{reviewquality.pdf}.
On average and
roughly speaking, one third of reviews is considered good,
one third reasonable, and one third either unhelpful (20\%) or
grossly faulty (10\%).
However, there is considerable diversity in the opinions:
The most optimistic quarter of respondents believes 50\% or more of
all reviews are good, 
while the most pessimistic quarter believes only 21\% or less are good.
(For comparison, the \citet{mulligan2013} survey found high or very high
review quality satisfaction for 69\% of respondents.)
% (We should not treat 'reasonable' as equivalent to 'high')

We perceive these responses as balanced (rather than cynical).
They do not paint a rosy picture of getting one's work reviewed in SE:
In a typical set of three reviews, one has to expect that only one of them
will be as thorough and helpful as they all should be, while the other two are
not. As a result, acceptance decisions will be highly noisy.

%------------------------------------------------------------------------
\subsection{Why are the faulty reviews faulty?}\label{faulty-why}
%columns AA-AC

We asked \wesay{In your opinion, what were the main reasons for unhelpful
  and/or grossly faulty reviews (if any)?} 
and received 136 answers.
In those, our open coding found 25 different reasons mentioned and 
276 mentions (100\%) overall.

Two reasons stand out:
% round(c(66, 60)/276*100)
Reviewers not allocating enough time (24\%) and
reviewers being insufficiently familiar with the topic of the work (22\%).
Some of these answers sounded cynical (even sarcastic) or sad, but
most were matter-of-fact; we made no attempt to code emotional quality.
A few answers captured a lot of their issue succinctly:
\theysay{lack of time or effort};  % line 22
\theysay{In cases it is simply because the reviewer did not do his/her
  job, or accepted to referee a paper for which he/she was not
  qualified. But when you submit to good venues, with good PCs, that
  happens less frequently.}.  % line 9

Six other reasons were mentioned at least a dozen times:
% round(c(27,18+4,14,14,13,12)/276*100)
The reviewer does not care to make a good review (10\%),
the reviewer is biased towards some type of research content or method (8\%),
misunderstandings (5\%),
generally low reviewing skill (5\%),
inappropriate priorities set by the reviewer (5\%), and
exaggerated expectations (4\%).

So at least one third of mentioned reasons (lack of time and
lack of care) ought to be repairable.
We are not aware of other studies asking respondents why faulty
reviews become faulty.

%------------------------------------------------------------------------
\subsection{Worst peer review experience as author}\label{worstexperienceauthor}
%columns AD-AF

We asked \wesay{As an author, what has been your worst
  experience with peer review?} 
and received 123 answers.
In those, our open coding found 37 types
of experience mentioned and 158 mentions (100\%) overall.

The most common topic was a lack of justification in a review:
Unjustified rejection (13\% of mentions),
% unjustifiedrejection, unjustifieddeskrejection
unjustified individual points of criticism (7\%), or
% unjustifiedcritique, unjustifieddoubt
a discrepancy between the decision and the review text (6\%).
% unjustifiedreview
The idea of our question was to collect anecdotes and indeed many
respondents provided such stories. 
Some of them included evidence that the issue with the reviews was not
merely imagined, like this one:
\theysay{A paper being rejected with very short reviews that gave no
  indication as to the reasoning behind the decision. While everyone
  has a horror story about a rejection, I had a paper that was
  submitted to a journal and rejected without review by the editor:  I
  submitted the paper to another journal and it was fast-tracked into
  the next available issue and now has over 300 citations (Google
  scholar).}
Or this one:
\theysay{I got one paper rejected because it ``didn't even cite
  [XYZ]''. The reviewer accused us of having no clue about the
  field and not knowing even the most elementary works in the
  field. Therefore, he refused to review the paper any further, i.e.,
  the review was just a couple of sentences long. Interestingly, one
  of the authors of [XYZ] [...] was also an author of the paper that
  got this crappy review. 
  We of course [were] fully aware of [XYZ], but did not find
  it relevant for what we presented.} 

Next in line (and in fact related) are 
\theysay{lazy reviewers} (9\%) and
overly short reviews (7\%).
Examples:
(1)~\theysay{The review was one line: You failed to convince me this is
  an interesting idea.},  % LICS conference
(2)~\theysay{A 10,000 word manuscript fetching a 250 word review, 
  out of which 200 words are spent in summarizing the manuscript},
(3)~\theysay{Reviewers [...] stop reading the paper after the
  abstract},
(4)~\theysay{Reviews which were not only misguided in their criticism, but
  entirely indecipherable due to 
  reviewers' evident off-the-cuff writing (fragmented sentences, lack of
  clarity, reference to misspelled terms). This is particularly galling
  because a.) it is not possible to extract valuable criticism when the
  reviewer seems not to have read the same paper as you wrote; b.) it is
  particularly insulting when a rejection is so obviously unconsidered
  that the reviewer hasn't read it to themselves.}.

6\% of mentions (that is, 7\% of respondents) state they never had
a particularly bad reviewing experience.
On the other hand, 4\% report direct insults or criticism addressing the
researcher rather than the work (argumentum ad hominem).
% insultingcomments, adhominemrejection
Examples:   % about OOPSLA
(1)~\theysay{At [XYZ], I have seen a considerable amount of things like name
  calling.  My students have been told they are schizophrenic, 
  directly in peer review, for the
  bizarre crime of running empirical studies and reporting the data.
  [...] I am consistently amazed at the total lack of
  accountability in reviewing -- even 3rd grade level name calling is,
  somehow, allowed in a venue like that.}.
% about ICSE:
(2)~\theysay{a reviewer made personal attacks on one of my co-authors.
  Fortunately, my co-author took it with good humor, but I
  felt that it was unacceptable.  I reported it to the PC chairs though
  there is no way to confirm that any word ever made it to the reviewer.
  I hope it was just a momentary lapse of judgment on the part of the
  reviewer, but it definitely reduced my opinion of [XYZ] as a venue
  considering that the PC chairs did not even acknowledge it.}.
Further anecdotes revolved around other types of suspected abuses of 
reviewer anonymity:
% about TOSEM
(3)~\theysay{A review process for [journal XYZ] where a
  well-established author on the same topic did not want a new actor
  around, pretending that everything was already done by him and his
  research group, which was clearly false.}.
(4)~\theysay{an expert disagreeing because the material presented showed
  that the results of one of their earlier articles were flawed.
  It makes no sense to suppress articles that attempt reproduction,
  certainly not articles that provide a detailed basis why earlier
  results are flawed. This is also a failure of the committee in
  general, for not seeing the conflict of interest.}.
Reviewer anonymity will be a topic of Section~\ref{blinding}.

Four other categories have 6 or 7 mentions (4\%) each.
They speak about 
unqualified reviewers, 
unconstructive criticism,
reviewers unjustifiedly pushing their own work, or
reviewers severely lacking an understanding for the nature of
empirical work.
Examples of the latter:
(1)~\theysay{I did not have 100\% response [rate in] a survey},
(2)~\theysay{Why did you not just measure and see what the result is?},
(3)~\theysay{Desk-rejected qualitative research: There are no numbers!}.

The rest is a long list of rare problems (just 1 or 2 mentions)
that includes anything 
from administrative problems
over a decision based on only a single review
to receiving a review that was obviously written for a different submission.
Several of these relate to various types of pickiness
and one of them is particularly worrying:
\theysay{I, sometimes, feel that if I'm too honest about the limitations of a
  technique, reviewers will simply pick on it. Of course, if the
  limitations are too large to render the technique useless, then I
  agree that it is a big issue. However, what I often find is that if
  I hadn't mentioned that limitation in the first place, the reviewers
  wouldn't have picked up on it.}

In order to end on a more positive note, we quote this participant:
\theysay{All this sounds like complaining, I am sure.
  I accept that the review process is a human process and therefore
  filled with problems.  But I hope to make it as good as possible.}

We are not aware of other studies asking respondents to openly report
their worst experiences with peer review; our results therefore  
complement quantitative results on preconceived problem areas such as
those provided by \citet{mulligan2013}. 
% our respondents have a conference mindset. Comparing their answers
% to Mulligan's journal-related ones is not sound.

%------------------------------------------------------------------------
\subsection{Worst peer review experience as reviewer}\label{worstexperiencereviewer}
%columns AG-AI
We also asked the same question from the other perspective:
\wesay{As a reviewer, what has been your worst experience with peer review?} 
and received 111 answers.
In those, our open coding found 53 different types
of experience mentioned and 142 mentions (100\%) overall.

The most frequent answer (17 times, 12\% of mentions) is that reviewers 
never had a \emph{particularly} bad experience, e.g.: 
\theysay{Not too many, typically I've met my
  peer-reviewers, so we all behave pretty civilized.}

Among the rest, four of the types stick out, 
at 11 to 13 mentions each (8\% to 9\%; 
the next-lower one has only 4\%).
At the top of the list are poor-quality submissions.
Here are some variants of that:
(1)~\theysay{Having to read papers which should have been desk-rejected as
  unreadable.},
(2)~\theysay{Articles that are so bad and unreadable that they are an insult
  to the time reviewers voluntarily and freely spend on this process.},
(3)~\theysay{Not particularly horrible, but
  once I had to review a paper that was too abstract and
  general. There was nothing that could be criticized about it, it was
  a vision paper and the authors were established members of the
  community. None of the three reviewers had anything to critique
  about the paper, but it was also clear that there was not much of
  substance in it. The paper ended up marked borderline/weak accept
  and eventually being published.}.

Second in line are authors that do not make the necessary improvements
to their article, as in these stories:
(1)~\theysay{Spending a lot of time on reviewing a conference paper
  and discovering conceptual flaws when such flaws are then met with 
  apathy by the authors and other reviewers. 
  It is sad when such flaws are not documented in the final
  paper version (or explained why they are in fact not flaws). To me
  this is a big drawback of the conference publication model, since a
  journal editor can enforce that authors respond to such criticism.},
(2)~\theysay{In a journal or conference with a revise-resubmit process, 
  I think finding authors who dismiss or ignore feedback is particularly
  insulting, especially if I've spent quite a bit of time to thoroughly
  read their work and think about my feedback.},
(3)~\theysay{I don't know whether [ignoring my improvement request] was 
  because he considered my request unreasonable, because he had lost the 
  raw data, or because [fixing the problem] showed that his results were 
  not very strong.  This is particularly bad because it shows that authors
  can selectively present data that strengthens their claims, and 
  the review process is not strong enough to guard against it},
(4)~\theysay{Journal A review: I recommended major revision, and 
  really major it would have to be. 
  My review contained about three dozen issues.  
  New version of the article comes in: the two smallest issues have
  been addressed, none of the important ones are.  
  I state this and reject the article. The editor rejects the article. 
  This was on a Thursday.  
  On the following Monday, journal B queries me for a review.
  It turns out it is the same article again, in exactly the version 
  rejected by journal A on Thursday.}
One of the respondents remarked on a similar story as follows:
\theysay{Hmm, is double-blind reviewing going to make such behavior
  more common? That would be horrible.}.

Third is what reviewers perceive as inappropriate behavior (including
passivity) of editors, PC chairs or other powers-that-be,
for example:
(1)~\theysay{Encouraging PC chairs [...] to get papers accepted,
  because the acceptance is too low.},
(2)~\theysay{The worst experience was at [XYZ], where several of the
  decisions PC members came up with after long and careful discussions have
  been overruled [...] without substantial arguments and without
  asking back.},
(3)~\theysay{[I rejected an article] that used students as subjects
  because it was in violation of the basic rules of ethics. The other
  reviewers accepted the paper on the grounds that the results were
  good, despite the fact that the non-compliance with ethical norms
  could have introduced serious threats to validity in the results.},
(4)~\theysay{I think physical PC meetings reward fast thinking and
  good communication skills, without analyzing in depth the issues in the paper.
  I think online discussions work better than PC meetings in this respect. 
  Moreover, PC meetings tend to be too far away from the time
  when papers are read and reviewed, especially if authors are granted a
  chance of rebuttal, which extends the reviewing time line.}.
On the other hand, remarks elsewhere show that reviewers may have more
influence than some of them may believe, like here:
(1)~\theysay{It took a lot of dialog with the
  editor to sway him/her from applying a simple vote.},
(2)~\theysay{I decided not to participate as a member of the program
  committee in the future.}.

Rank 4 belongs to the first of many types that echo all of the issues
brought up from the author perspective, just this time the reviewers
criticize their co-reviewers (and sometimes themselves). 
For example:
(1)~\theysay{The other three reviewers wrote bland, nothing-type reviews and
  were impressed by a lot of statistical mumbo-jumbo which was actually
  badly flawed. The paper claimed to be a how to do it type article and
  therefore particularly dangerous.  It took a lot of dialog with the
  editor to sway him/her from applying a simple vote.},
(2)~\theysay{Papers from a completely unfamiliar area where I could not
  validate the results or determine their significance.}.
We split the bad co-reviewers issues into many types, such as
the reviewer being lazy, dogmatic, inflexible or
the review being too short, unbalanced, unjustified, too critical, uncritical.
Had we collected all of these under a single type, it would have
ranked at the top by far with 28 mentions (20\%).

Some of the remaining (rare!) issues concern cases of power abuse on
the side of reviewers or editors. Examples are:
(1)~\theysay{We once had a reviewer from
  [country XYZ] on our PC who would give all papers from [country XYZ]
  the very best grades, even if everybody else did not like the paper. 
  He then did not even show up at the PC meeting.  
  All his reviews were canceled, and we PC members were in for a night shift.},
(2)~\theysay{Reviewers pushing papers of well known authors (or authors who
  are their friends) to get the paper accepted.}
(3)~\theysay{Seeing other reviewers writing just 1 or 2 line
  without saying anything on the paper (and probably not reading it),
  and still fighting to accept/reject the paper.}.

However, please remember that the most frequent reply type was the
no-major-problems type, e.g. \theysay{Not much.}

Similarly to Section \ref{worstexperienceauthor}, 
we are not aware of other studies asking respondents to openly report
their worst experiences with peer review as reviewers.
Our open-ended exploration offers insights that previous quantitative studies
have not provided.

%========================================================================

\section{Results: How much blinding is appropriate?}\label{blinding}

Background: ICSE has traditionally used a single-blind reviewing
regime; 
ICSE 2018 is the first to switch to double-blind reviewing.
Past ICSEs (to our knowledge, from 1987 to 1991) have required that one of the reviewers be listed at the 
foot of the title page of accepted articles as having ``recommended''
the work.

\Rplotw{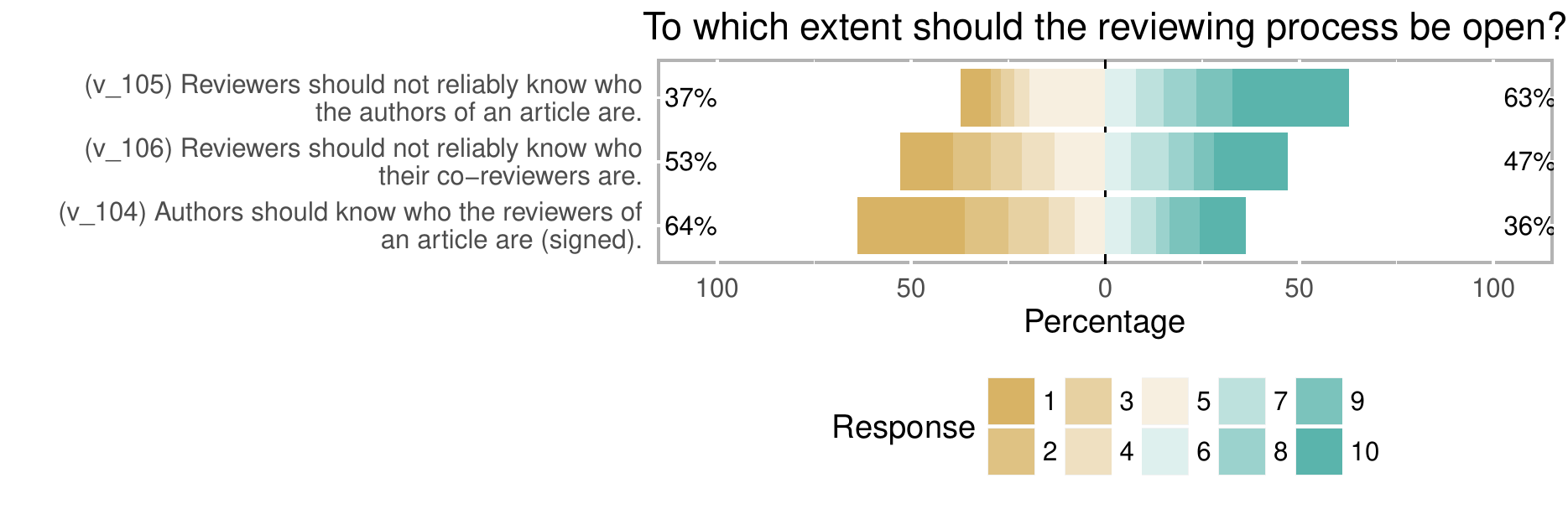}{How much blinding is appropriate?
  Answers range from strong disagreement (1) to strong agreement (10).}{0.9}

We had three agree/disagree items on this topic, which all received
$n=160$ answers (100\%).
The items' wording and the response percentages are shown in 
Figure~\ref{reviewopenness.pdf}

There is a two-thirds majority agreeing that reviewers should not know
author names (in practice, this means the ``double-blind'' regime).
In \citet{mulligan2013}, 76\% of their cross-discipline participants considered
double-blind peer review the most effective method;
in \cite{Bacchelli2017}, 46\% of their software engineering
participants were in favor
of all software engineering venues to go double-blind.

As for blinding reviewers with respect to the names of the
co-reviewers, sometimes called ``triple-blind'', respondents are split
half-and-half.

A potentially surprising result arises for the third question, zero-blinding:
About one third of respondents say they believe reviewers ought to give up 
their anonymity and sign their reviews.
The OpenAIRE survey had not asked a ``should'' question, but even for its
very openness-minded participants only 44\% had agreed zero-blinding
would improve review quality \citep{RosDepSch17}.

%========================================================================
\section{Results: Should drafts and reviews be laid open?}
\label{openness}

\Rplotw{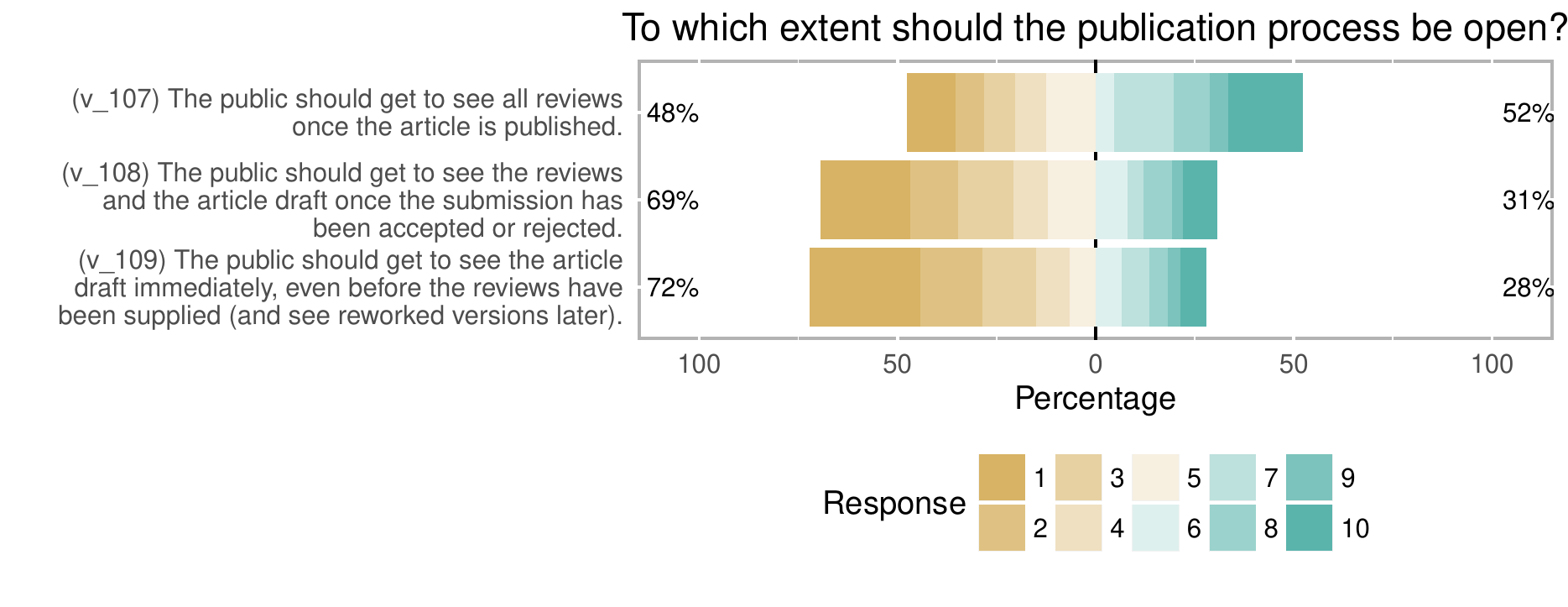}{Should drafts and reviews be laid open?}{0.9}

We had three agree/disagree items on this topic, which also all received
$n=160$ answers (100\%).
Their wording and the response percentages are shown in 
Figure~\ref{publicationopenness.pdf}.

Respondents are split half-and-half about whether reviews should be
published along with an accepted article.
There is also some limited support for the more radical ideas of
publishing article draft and reviews also for rejected articles
(31\%),
or even publishing the article draft immediately upon submission (28\%).
These sentiments are more positive than those found by \citet{mulligan2013} and
only moderately less enthusiastic than those in the OpenAIRE survey
\citep{RosDepSch17}.

%========================================================================
\section{Results: Should reviewers be compensated?}
\label{compensation}

%------------------------------------------------------------------------
\subsection{Monetary or quasi-monetary compensation}\label{quasimonetarycompensation}
%v_55 yes/no, v_56 comment

We asked 
\wesay{Reviewers should receive a (quasi-)monetary compensation
  for their work (e.g. memberships, subscriptions, registration
  discounts, or money payment). If so, which?}.
41\% of $n=160$ respondents agreed (to varying degrees) and
33\% provided a free-text comment on the issue.
In those, our open coding found 14 different types
of suggestion mentioned and 142 mentions (100\%) overall.
16\% of those suggest monetary compensation, 
nearly all of the others suggest variants of the other ideas mentioned
in the question, the most popular being 
conference registration discounts (43\%) and waivers on 
society memberships (7\%) or subscriptions (10\%).

These results reflect a much more honor-based attitude towards
reviewing than those from \cite{mulligan2013}, where 41\% of
participants showed inclination towards monetary and 51\% towards
quasi-monetary compensations.
% our 16% monetary mentions represent much less than 16% of participants!

In contrast, a few of our respondents even made critical remarks on the
for-profit culture
in much of the scientific publishing system, like this one:
\theysay{It is ridiculous we are doing free work that will ultimately result
  in more money for Elsevier/IEEE. Willing to debate if money should
  go to person or institution}.

%------------------------------------------------------------------------
\subsection{Showable proof of good work}\label{nonmonetarycompensation}
%v_58 yes/no, v_59 comment

We also asked 
\wesay{Reviewers should receive showable proof for good reviewing work
  (e.g. public visibility of their review texts, or a reviewing quality
  certificate). If so, which?}.
71\% of again $n=160$ respondents agreed and 51\%
% 113 yes, 81 with comment
provided a free-text comment on the issue.
As for the monetary compensation question, these comments were heavily
primed by the examples given in the question,
but contain a number of further ideas as well.
In the 81 comments, our open coding found 31 different types
of suggestion mentioned and 116 mentions (100\%) overall.

The most common suggestion was indeed handing out a certificate (suggested by 
%26 + 16 + 13 + 2 + 1 + 1 = 59
51\% of mentions). 
Some of these were more specific, for example they proposed that
only the best reviewers should get a certificate (16\%)  % 18
or the certificate should state the quality of the reviews (11\%). % 13
Various ideas amounting to other forms of transparency or a reputation
system, when taken together, represent another 27\% of mentions,
so that these two categories sum to 78\% of mentions, making
everything else minor.
Interesting specialized points made by only one or two respondents in
these two or other categories include:
publish certificates centrally, use Publons, mimic StackOverflow,
don't forget the subreviewers, blacklist bad reviewers.

The only \citet{mulligan2013} equivalent is ``Acknowledgment in the
journal'', which 40\% of respondents found attractive.
The OpenAIRE survey does not report on this issue.
\citet{warne2016} is specifically about compensation and reports mean
agreement of 4.2 (on a 1-to-5 scale) with ``Reviewing should be
acknowledged as a measurable research output''.

Summing up, there is a lot of support for issuing some kind of
reviewing certificate to reviewers and
some support for various forms of quasi-monetary compensation.
Overall, our respondents are far more welcoming to 
showable proof (71\% agreement) than to 
monetary or quasi-monetary compensation (41\% agreement).

Three initiatives are already pursuing goals of the ``showable proof'' type
on a general level:
Publons\footnote{\url{http://publons.com}} (for journals only)
counts reviews and also allows publishing them, 
Academic Karma\footnote{\url{http://academickarma.org}} 
aims at making the content of all
reviews and review responses public and allows signing reviews,
Review Quality Collector\footnote{\url{http://reviewqualitycollector.org}} 
(RQC, currently for conferences only, later also for journals)
issues certificates based on an explicit quantitative
review quality assessment.

%========================================================================
\section{Results: How should the reviewing regime change?}\label{changes}

At the end of our survey, we asked our respondents 
\wesay{If you could change the current review practices at will,
what would you consider the most valuable improvements (and why)?}.
This was a free-text question only (no predefined categories at all)
and it received 118 responses.
In those, our (somewhat over-eager) open coding found an enormous 57
different types
of change suggestion mentioned with 162 mentions (100\%) overall.

Among these, two stick out by a far margin (with 17\% and 15\% of
mentions, respectively): introduce double-blind reviewing
and introduce open reviewing.
Open reviewing in this sense is a combination of publishing the
review (and perhaps the draft submission) and 
attaching the reviewer's name to it, but the respondents
provided very different amounts of detail in their description so not
all of them may actually have meant all of these elements. 
One could actually use both suggestions in one process: prepare the
initial review under a double-blind regime, perhaps even have a
discussion with the authors still in double-blind mode, and then
lift anonymity on both sides and publish the reviews (for
accepted or all submissions) and perhaps the article drafts as well.
5 people indeed mentioned both together, which is logical if one
subscribes to \theysay{The most important thing is
  to have a symmetrical reviewing process (i.e., either blinded or
  unblinded).}.
Most proponents of one of these ideas, however, do not favor the
other,
with attitudes like this one for the open reviewing camp:
\theysay{So many [reviews] are so insightful, everyone should be able
  to learn from their critiques!} 
or this one against:
\theysay{Making reviews more visible opens up a can of worms that
  will, ultimately, not be helpful. People are vindictive, if you
  haven't noticed. Even the best researchers can have their moments.}
and this one for the double-blind camp:
\theysay{There is evidence of bias in scientific reviewing and
  evidence that double-blind reviewing can reduce it.  Experience with
  light double blind reviewing in related fields suggests that it is
  successful, low-cost, and has few drawbacks.}
or this one wary of it:
\theysay{Double blind has good justifications, but it is unfortunate
  it also lowers ability of established researchers to push the envelope: 
  the low accept rate makes blinded highly novel papers have low
  chance of acceptance without a level of experimental evidence [...]}.

Following these top suggestions are three with 4\% of mentions each:
reward reviewers, rate reviewers, decrease reviewing load.
Also with 4\% of mentions comes a category 'novel process' with
sketches of radical ideas such as this one:
\theysay{Reviewers rank all papers. Authors decide whether or not they
  accept to present their work. If and only if they present their
  work then their rank is published with a review summary. Author
  presentation time is proportional to their rank position.} 

The long tail contains a number of straightforward suggestions such as
reducing reviewing load, avoiding sub-reviewers, or introducing rebuttals
as well as a few more far-reaching suggestions
we found remarkable:
(1)~Elect (rather than appoint) program committees.
(2)~\theysay{Build some sort of reviewer rating system. 
  Reward good reviewers and warn and eventually punish bad reviewers.
  Build a culture of valuing good strong reviews.}, which becomes most
interesting in combination with \theysay{People who publish
  (including co-authors) should be required to review a comparable amount.}.
(3)~Get rid of publishing papers at conferences: \theysay{In Computer Science,
  [we should move] away from a model in which excellent research is
  routinely rejected only because there was another paper at the same
  conference that the PC considered to be of equal quality but higher
  excitement level}.
(4)~Physical PC meetings are an important practice for some 
  (\theysay{there are several best practices that help maintaining high
  standards, most notably physical PC meetings}) and
a threat to good peer review for others (\theysay{they are a drain on 
  everyone and on the environment, and they tend to favor outcomes
  advocated by strongly vocal members.}.

We are not aware of similar results from other studies.

%========================================================================
\section{Results: How these results may change in the future}
\label{future}
\label{resultslast}

The responses to our questions regarding blinding, publishing reviews
and drafts, or compensating reviewers represent attitudes.
How will these attitudes change in the future?
We asked our respondents for their age, so we can (and now will)
look for age-related trends in our data.

%------------------------------------------------------------------------
\subsection{Theoretical assumptions, approach}

If we set aside the case of disruptive changes and look only at trends already
represented in our dataset, we see two possibilities:
\begin{itemize}
\item Hypothesis G: There is a generational trend;
  the attitude of a person is largely stable.
  If only G were true, the same attitude of our now-younger respondents
  today would largely be the attitude of then-older respondents in twenty years.
\item Hypothesis S: There is a seniority trend;
  the attitude of a person changes with experience and/or role.
  If only S were true, the attitude of our now-older respondents today
  would likely be the attitude of then-older respondents in twenty
  years as well.
\end{itemize}
Obviously, we should expect a mix of both effects.
But is one of them dominant?
Our data cannot provide a definite answer, but can provide a strong
clue, because we have a good proxy for seniority, experience, and role 
in our data:
The current professional position a respondent holds.

We will build linear models of attitudes using age and seniority as
predictors and see whether one or both are statistically significant
and how large their coefficient (i.e., the respective effect) is.
Where only age is significant, this indicates G is dominant.
Where only seniority is significant, this indicates S is dominant.
Where both are significant, this indicates both effects mix.
Where none of them is significant, this indicates time trends are weak
or non-existing.

There is a problem: age and seniority correlate strongly.
Therefore, we should not expect the linear models to be highly
stable\footnote{The instability induced by predictor collinearity is
  commonly measured by the variance inflation factor (VIF). 
  Values under 4 are generally considered totally unproblematic
  \citep{OBrien07}.
  The VIFs in our models range vom 1.39 to 1.82.}
% > print(vif(lm(openness ~ ageD + prof, data=df)))
%     ageD     prof 
% 1.387703 1.387703 
% > print(vif(lm(openness ~ ageD + tenured, data=df)))
%     ageD  tenured 
% 1.820827 1.820827 
Therefore, we will only be able to say which of G or S is dominant if the
difference between the two effects is large.

%------------------------------------------------------------------------
\subsection{Predictor variables}

We will use the following predictor variables in the models:
\begin{itemize}
\item \textbf{ageD}: age in decades. We use decades rather than years
  to make the coefficients larger and easier to read.
\item \textbf{prof}: whether or not the respondent is a tenured or
  non-tenured professor; this is a proxy of seniority.
\item \textbf{tenured}: whether or not the respondent is a tenured
  professor; this is an alternative proxy of seniority.
\end{itemize}
Each model will have age as a predictor plus zero or one of the
seniority measures, plus possibly the interaction of age and
seniority.
For the latter, we will use the notation of R in the coefficient
table, e.g. ageD:profFALSE and ageD:profTRUE.
prof and tenured, being only binary, tend to have less predictive
power, giving the G effect a head start, which we need to keep in mind
for the discussion.

%------------------------------------------------------------------------
\subsection{Dependent variables}

We try each of the following dependent variables in the models:
\begin{itemize}
\item \textbf{Pgood}: the percentage of ``good'' reviews.
\item \textbf{Punhelpful}: the percentage of ``unhelpful'' reviews. 
  (The ``reasonable'' ones appear less interesting.)
\item \textbf{Pfaulty}: the percentage of ``grossly faulty'' reviews.
\item \textbf{AknowR}: whether authors should know who their reviewers are. 
  This, as all of the
  other attitude variables below, is measured on the 10-point
  disagree/agree scale which we always interpret as a difference scale here
  and represent it by evenly-spaced fractional numbers in range $-5\ldots 5$.
\item \textbf{RknowA}: whether reviewers should know who their
  authors are.
\item \textbf{RknowR}: whether reviewers should know who their
  co-reviewers are.
\item \textbf{openness}: the average of the above three.
\item \textbf{opennessChg}: ditto, but with the sign of the latter two
  components reversed. 
  This represents the attitude towards change relative to the
  single-blind regime that was most common in software engineering
  (in particular: used at ICSE) in the timeframe we asked about.
\item \textbf{pubreviews}: whether or not reviews should be published
  along with accepted articles.
\item \textbf{publicness}: the average of all three publicness-related
  questions we asked.\footnote{We do not use the other two separately because
  they were formulated in a manner that makes their solo
  interpretation ambiguous.}
\item \textbf{monetary}: whether or not reviewers should receive
  monetary compensation for their work.
\item \textbf{certificate}: whether or not reviewers should receive
  ``showable proof'' of good reviewing work.
\end{itemize}

%------------------------------------------------------------------------
\subsection{Model selection method}\label{modelselection}

For each of the 12 dependent variables, we will consider five different models 
as follows (60 different models overall) and present only the most
convincing one from each block -- or none, if none is convincing at all.
A convincing model needs all coefficients to have statistical
significance and a high $R^2$.
Each candidate model has theoretical plausibility, so we do not
consider this procedure to constitute ``fishing for significance''. 
Nevertheless, we will use a low significance threshold of $p < 0.02$
for each coefficient to reduce false positives.

We will show this procedure by spelling the process out for one of the
dependent variables; we will only present end results for the
remainder.
Consider Table~\ref{modelsexample}:
Each block of rows represents one model, numbered in the leftmost
column.
The second column describes the predictors used in the model:
age and prof separately (A+P);
age-and-prof interaction (A:P);
ditto for tenured (A+T; A:T); or
age only (A).
The third and fourth column show the coefficients in the model;
the fifth the corresponding $p$ value;
the final column shows adjusted $R^2$ for the model: The fraction of
variance explained after deducting the random-chance component for
each degree of freedom used by the model.

\begin{table}
\centering
\caption{All five candidate models for dependent variable opennessChg.
  Our model selection rules suggest to use model 40 for this variable.
  Columns and model selection are explained in the main
  text.}\label{modelsexample}
\begin{tabular}{rclrrr}
 &type&name&coeff.&\multicolumn{1}{c}{$p$}&\multicolumn{1}{c}{$R^2$}\\ \hline
36& A+P  & (intercept)        & 2.48    & 0.001 & 0.070 \\
  &      & ageD               & -0.60   & 0.003 &       \\
  &      & profTRUE           &  0.03   & 0.950 &       \\ \hline
37& A:P  & (intercept)        & 2.50    & 0.003 & 0.070 \\
  &      & ageD:profFALSE     & -0.60   & 0.018 &       \\
  &      & ageD:profTRUE      & -0.60   & 0.002 &       \\ \hline
38& A+T  & (intercept)        & 2.47    & 0.002 & 0.070 \\
  &      & ageD               & -0.59   & 0.009 &       \\
  &      & tenuredTRUE        &  0.01   & 0.981 &       \\ \hline
39& A:T  & (intercept)        & 2.61    & 0.006 & 0.070 \\
  &      & ageD:tenuredFALSE  & -0.64   & 0.024 &       \\
  &      & ageD:tenuredTRUE   & -0.61   & 0.003 &       \\ \hline
40& A    & (intercept)        & 2.46    & 0.000 & 0.077 \\
  &      & ageD               & -0.59   & 0.001 &       \\ \hline
\end{tabular}
\end{table}

Model 36 (this will be the model number in our overall models list)
in Table~\ref{modelsexample} tells us that agreement with
reviewing regime change (opennessChg) is 2.48, halfway between total
agreement and a neutral stance, if the respondent is a baby (0 decades
old) and not a professor (profTRUE is 0). 
By the age of 20, agreement will have fallen
to 1.28 and by the age of 50 to -0.52.
However, the seniority effect (profTRUE) has has a non-significant
coefficient, so the whole model is not meaningful.
Model 38, which replaces prof by tenured, is very similar; 
both cannot be used.

Model 37 uses the \emph{interaction} of ageD and prof instead of using
them side-by-side; all three
coefficients are significant, so this could be a useful model.
However, both coefficients of the interaction are practically the
same, so this is not a meaningful model either.
Its cousin, model 39, behaves similarly.
The coefficients are a bit more different, but the first interaction coefficient
is no longer significant, so this model also cannot be used.

Model 40, the simplest of them all, using only age for prediction, is
the only convincing model and hence the one we select.
Much like model 36, it says agreement is at 1.28 for the average 20-year-old
respondent and -0.49 for a 50-year-old.

%------------------------------------------------------------------------
\subsection{Results}

\begin{table}
\caption{Best model for each dependent variable where a convincing
  model was found.
  The interpretation is explained in the main text.}\label{models}
\tabcolsep 4pt
\centering
\newcommand{\N}[1]{\multicolumn{2}{l}{#1}}
\begin{tabular}{rclrrr}
 &type&name&coeff.&\multicolumn{1}{c}{$p$}&\multicolumn{1}{c}{$R^2$}\\ \hline
25& A:P  & (intercept)        &-3.93    & 0.000 & 0.028 \\
\N{RknowA}& ageD              & 0.61    &(0.026)&       \\ \hline
30& A    & (intercept)        &-3.05    & 0.011 & 0.041 \\
\N{RknowR}& ageD              & 0.76    & 0.009 &       \\ \hline
40& A    & (intercept)        & 2.46    & 0.000 & 0.077 \\
\N{opennessChg}& ageD         & -0.59   & 0.001 &       \\ \hline
52& A:P  & (intercept)        & 0.99    & 0.000 & 0.050 \\
\N{monetary}& ageD:profFALSE  & -0.17   & 0.007 &       \\
  &      & ageD:profTRUE      & -0.14   & 0.003 &       \\ \hline
57& A:P  & (intercept)        & 1.09    & 0.000 & 0.015 \\
\N{certificate}&ageD:profFALSE& -0.11   &(0.046)&       \\
  &      & ageD:profTRUE      & -0.07   &(0.072)&       \\ \hline

%% 11& A+P  & (intercept)        & 2.00    & 0.000 & 0.000 \\
%%   &      & ageD               & -0.60   & 0.000 &       \\
%%   &      & profTRUE           & -0.01   & 0.000 &       \\ \hline
%% 12& A:P  & (intercept)        & 2.00    & 0.000 & 0.000 \\
%%   &      & ageD:profFALSE     & -0.60   & 0.000 &       \\
%%   &      & ageD:profTRUE      & -0.60   & 0.000 &       \\ \hline
%% 13& A+T  & (intercept)        & 2.00    & 0.000 & 0.000 \\
%%   &      & ageD               & -0.60   & 0.000 &       \\
%%   &      & tenuredTRUE        & -0.01   & 0.000 &       \\ \hline
%% 14& A:T  & (intercept)        & 2.00    & 0.000 & 0.000 \\
%%   &      & ageD:tenuredFALSE  & -0.60   & 0.000 &       \\
%%   &      & ageD:tenuredTRUE   & -0.60   & 0.000 &       \\ \hline
%% 15& A    & (intercept)        & 2.00    & 0.000 & 0.000 \\
%%   &      & ageD               & -0.60   & 0.000 &       \\ \hline

\end{tabular}
\end{table}

We discuss all dependent variables in order.
Where a convincing (or semi-convincing) model was found, it is shown
in Table~\ref{models}.

%### Pgood, Punhelpful, Pfaulty:
Models 1--15.
Neither of the variables Pgood, Punhelpful, Pfaulty has any model at
all with all-significant coefficients, so models 1 to 15 are all
missing from the table.
This tells us that the perception of review quality appears to be a
timeless phenomenon.\footnote{More precisely: Age and seniority
  effects are too weak to show up in a dataset of the size we have.
  This also suggests the estimates are largely unbiased.} 
We will use this term, timeless, for similar cases of
no-good-model-at-all below.

%### AknowR, RknowA, RknowR:
Models 16--30.
AknowR, the practice of reviewers signing their reviews, is timeless: 
Our respondents are all similarly skeptical.
Strictly speaking, RknowA is timeless as well, but it has one model, 25,
close enough to significance that we include it here for information:
Older respondents appear much less adamant that authors should be
hidden from reviewers.
RknowR model 30 shows a relatively strong age effect: 
Younger respondents tend to prefer
hiding reviewers' names from each other, ones over the age of 40 no
longer think so.

%### openness:
Models 31--35.
AknowR, RknowA, RknowR all represent a form of openness (transparency)
in the reviewing process, so averaging them describes attitudes towards
openness in general. But none of the models 31 to 35 is convincing;
the openness attitude as a whole is timeless.

%### opennessChg:
Models 36--40.
Given that it is currently the norm in software engineering (at
conferences) that reviewers know authors and each other, we
can reverse the sign of variables RknowA and RknowR and compute an
``inclination towards change from the current openness regime''.
This is the variable we have discussed in the example in
Section~\ref{modelselection}:
Young respondents are inclined to change, older ones much less so
(model 40).

%### pubreviews, publicness:
Models 41--50.
How about some other form of transparency: publishing the reviews?
Both variables, pubreviews and publicness, are completely timeless:
None of the models can explain more than 0.4\% of the variance.

%### monetary, certificate:
Models 51--60.
Finally, there is the issue of compensating reviewers.
Model 52 is the first  two-variable model that comes out as most convincing.
It states that young respondents tend to think (if only with a rather
weak majority) that reviewers should be compensated (quasi)monetarily.
Older ones believe this less, reaching the zero point at the age
of 59 if they are not professors and 73 if they are.\footnote{The more
  exact coefficients are 0.9947, 0.1674, and 0.1361.}
The corresponding model 57 for \emph{non}-monetary compensation behaves in
the same manner, except its coefficients are not fully significant and the
age effect becomes so weak that only 100-year-olds\footnote{Professors
  even need to wait until they are 153.} stop
believing-at-least-a-bit that issuing certificates to reviewers would
be worthwhile.
We include this model for comparison.

%------------------------------------------------------------------------
\subsection{Interpretation}

Overall, there is 
not a single A+P model with a significant profTRUE coefficient and 
not a single A+T model with a significant tenuredTRUE coefficient.
This tells us that seniority effects, if they exist at all, cannot be strong.
Therefore, the age effects found are likely mostly 
generation effects (hypothesis G), not seniority effects (hypothesis S).

Summing the results up, we see that while there are generational
effects here and there, they are not very strong.
We should not expect reviewing to change drastically just because the
now-younger generation will advance to positions of power.
Explicit change initiatives will likely be required instead.

%========================================================================
\section{Limitations}
\label{limitations}

Despite our relatively good response rate of 29\%, our sample is
obviously not representative of the base population, as is easy to see
from the demographics in Section~\ref{demographics}.
By assuming age and status distributions for our base population we could in
principle correct for this distortion, but we consider this too
unreliable and so do not do it.
So the study is limited in that our characterization of what
population it represents remains imprecise.

As with any survey, the truthfulness and well-reflectiveness of the
answers is not certain, but we saw no signs of distorted responses
and our base population can be considered as a serious one. 
Therefore, we expect this problem to be negligible. 
The same is true for accidentally wrong inputs.

The evaluation, both the statistical and the qualitative one, is
mostly straightforward, so we do not expect grave mistakes to have
gotten seriously in the way of the correctness of our reported results. 
In any case, other researchers can check based on our fully disclosed data (see Section ~\ref{sec:online-material}).

The strongest threat to validity concerns Section~\ref{future}:
Our trend analysis requires the assumption that attitudes such as those
surveyed here tend to be stable.
There is evidence that this is the case \citep{Schwarz2001},
but it still remains an assumption.
Fortunately, no strong conclusions arise from that analysis and need
to rest on that assumption, so the
actual threat to validity is small.

%========================================================================
\section{Conclusions}
\label{conclusions}

Our survey of perceptions of and attitudes towards contemporary peer
review in software engineering research among a rather senior subset
of the ICSE 2014, 2015, 2016 reviewers and authors
brought the following major findings:
\begin{compactenum}
\item The respondents agree with a multitude of purposes that could be
  ascribed to peer review.
  The strongest agreement (at 96\%) is with the purpose of ensuring 
  the validity of the research in question
  (Section~\ref{purpose}).
\item The respondents are skeptical regarding the quality that
  software engineering reviews typically have today:
  On average, they deem only one third of all reviews they receive to
  be of good quality, while another third is either useless or grossly faulty
  (Section~\ref{how-well-quantitative}).
  If these perceptions are correct, software engineering reviewing 
  is severely broken
  and a lot of time, nerves, and goodwill of all involved get wasted.
  Given the key role of peer review in the scientific process,
  we should make efforts to improve this situation.
\item When asked for the reasons why the grossly faulty reviews are
  faulty, the respondents offered several dozen possibilities.
  The top three of these, however, cover half of the answers:
  Reviewers not investing enough time (24\% of mentions),
  reviewers not knowing enough about the subject area (22\%),
  and reviewers not caring to prepare a good review (10\%)
  (Section~\ref{faulty-why}).
  None of the three is insurmountable:
  Time investment is a matter of priorities,
  lack of expertise can be accommodated by not taking on the review in
  the first place, and
  a lack of care stems from a modifiable (if not easily) attitude.
  Improvement efforts can be successful in principle.
\item We asked respondents for their worst peer review experience from
  the author perspective, which sheds some light on how reviews are broken
  when they are considered to be very broken.
  The top few categories account for 42\% of mentions:
  A lack of justification of something important (26\%),
  overly short reviews (7\%) or 
  some other form of apparent reviewer laziness (9\%)
  (Section~\ref{worstexperienceauthor}).
  Even if many reviewers indeed lack the skill to notice which
  statements in their reviews require justification, this is something
  that could be taught and trained; the other problems are even
  more straightforward to avoid -- again: in principle; if one wants to.
\item Our question on the worst peer review experience from
  the reviewer perspective shows that authors and editors or PC chairs
  also have their share of responsibility for the bad state of peer review,
  but the manners and reasons are more varied here
  (Section~\ref{worstexperiencereviewer}).
\item As for blinding, two thirds of respondents agreed author names
  should be hidden from reviewers (double-blind reviewing),
  half agreed co-reviewer names should be hidden, and
  one third agreed reviewer names should \emph{not} be hidden from
  authors (zero-blind reviewing).
  So, although few software engineering reviewers currently
  appear to sign
  their reviews, not so few appear to be willing to go the
  route opposite to the current trend towards double-blind reviewing
  and go fully transparent instead
  (Section~\ref{blinding}).
\item Half of the respondents also agree that review texts should be
  published along accepted articles
  (Section~\ref{openness}).
\item 41\% of our respondents agreed that reviewers should receive
  monetary or quasi-monetary compensation for their work, 
  the latter being preferred
  (Section~\ref{quasimonetarycompensation}). 
\item 71\% agreed that reviewers should receive showable proof of good
  work as a compensation. 
  Half of the specific suggestions in this regard amount to some kind
  of certificate (Section~\ref{nonmonetarycompensation}).
\item When asked how they would change the current reviewing regime if
  they could, our respondents produced a broad set of suggestions.
  Two of them were  more popular than the rest:
  Introducing double-blind reviewing, covering 17\% of mentions, or 
  introducing open (zero-blind and published) reviewing at 15\%.
  (Section~\ref{changes}).
\item We investigated how many of the above estimates and attitudes
  depend on age and/or seniority (i.e. experience and role)
  and found almost no seniority effects and
  only a few weak or modest age effects
  (Section~\ref{future}).
  This means one should not expect the reviewing regime to change
  quickly just because the current senior researcher generation retires;
  an explicit effort to change the attitude of the software
  engineering community as a whole will likely be required.
\end{compactenum}

\noindent 
Existing (non-SE) venues such as F1000Research, ScienceOpen, or The BMJ
show that ``radical'' solutions like non-anonymous, public reviewing
are possible;
many of the current issues with peer review quality likely shrink to
modest proportions under such conditions.
Furthermore, initiatives such as Publons and Review Quality Collector
provide ideas how even anonymous regimes can be improved.
How about some experimentation?

%========================================================================
\section*{Acknowledgments}

We thank our survey respondents for their sometimes quite elaborate, yet helpful
answers.
We thank all reviewers involved indirectly for participating in the
respective reviewing processes.
We also thank the anoymous reviewers of the present article, in particular
the one who semi-identified him- or herself as a member of the 
ICSE steering committee.
Daniel Graziotin has been supported by the
Alexander von Humboldt (AvH) Foundation.

\bibliographystyle{elsarticle-harv}
\bibliography{related}

\begin{thebibliography}{26}
\expandafter\ifx\csname natexlab\endcsname\relax\def\natexlab#1{#1}\fi
\providecommand{\url}[1]{\texttt{#1}}
\providecommand{\href}[2]{#2}
\providecommand{\path}[1]{#1}
\providecommand{\DOIprefix}{doi:}
\providecommand{\ArXivprefix}{arXiv:}
\providecommand{\URLprefix}{URL: }
\providecommand{\Pubmedprefix}{pmid:}
\providecommand{\doi}[1]{\href{http://dx.doi.org/#1}{\path{#1}}}
\providecommand{\Pubmed}[1]{\href{pmid:#1}{\path{#1}}}
\providecommand{\bibinfo}[2]{#2}
\ifx\xfnm\relax \def\xfnm[#1]{\unskip,\space#1}\fi
%Type = Article
\bibitem[{Armstrong(1997)}]{Armstrong1997}
\bibinfo{author}{Armstrong, J.S.}, \bibinfo{year}{1997}.
\newblock \bibinfo{title}{Peer review for journals: Evidence on quality
  control, fairness, and innovation}.
\newblock \bibinfo{journal}{Science and Engineering Ethics}
  \bibinfo{volume}{3}, \bibinfo{pages}{63–84}.
\newblock \DOIprefix\doi{10.1007/s11948-997-0017-3}.
%Type = Inproceedings
\bibitem[{Bacchelli and Beller(2017)}]{Bacchelli2017}
\bibinfo{author}{Bacchelli, A.}, \bibinfo{author}{Beller, M.},
  \bibinfo{year}{2017}.
\newblock \bibinfo{title}{Double-blind review in software engineering venues:
  The community's perspective}, in: \bibinfo{booktitle}{Proceedings of the 39th
  International Conference on Software Engineering Companion},
  \bibinfo{publisher}{IEEE Press}, \bibinfo{address}{Piscataway, NJ, USA}. pp.
  \bibinfo{pages}{385--396}.
\newblock \DOIprefix\doi{10.1109/ICSE-C.2017.49}.
%Type = Article
\bibitem[{Baxt et~al.(1998)Baxt, Waeckerle, Berlin and Callaham}]{Baxt1998}
\bibinfo{author}{Baxt, W.G.}, \bibinfo{author}{Waeckerle, J.F.},
  \bibinfo{author}{Berlin, J.A.}, \bibinfo{author}{Callaham, M.L.},
  \bibinfo{year}{1998}.
\newblock \bibinfo{title}{Who reviews the reviewers? feasibility of using a
  fictitious manuscript to evaluate peer reviewer performance}.
\newblock \bibinfo{journal}{Annals of Emergency Medicine} \bibinfo{volume}{32},
  \bibinfo{pages}{310–317}.
\newblock \DOIprefix\doi{10.1016/S0196-0644(98)70006-X}.
%Type = Misc
\bibitem[{BMJ(2017)}]{BMJopen}
\bibinfo{author}{BMJ, T.}, \bibinfo{year}{2017}.
\newblock \bibinfo{title}{Open peer review}.
\newblock
  \bibinfo{howpublished}{\url{http://www.bmj.com/about-bmj/resources-reviewers}}.
\newblock \bibinfo{note}{As of 2017-05-02}.
%Type = Article
\bibitem[{Bornmann and Mutz(2015)}]{bornmann2015}
\bibinfo{author}{Bornmann, L.}, \bibinfo{author}{Mutz, R.},
  \bibinfo{year}{2015}.
\newblock \bibinfo{title}{Growth rates of modern science: A bibliometric
  analysis based on the number of publications and cited references: Growth
  rates of modern science: A bibliometric analysis based on the number of
  publications and cited references}.
\newblock \bibinfo{journal}{J Assn Inf Sci Tec} \bibinfo{volume}{66},
  \bibinfo{pages}{2215--2222}.
\newblock \DOIprefix\doi{10.1002/asi.23329}.
%Type = Article
\bibitem[{Breuning et~al.(2015)Breuning, Backstrom, Brannon, Gross and
  Widmeier}]{breuning2015}
\bibinfo{author}{Breuning, M.}, \bibinfo{author}{Backstrom, J.},
  \bibinfo{author}{Brannon, J.}, \bibinfo{author}{Gross, B.I.},
  \bibinfo{author}{Widmeier, M.}, \bibinfo{year}{2015}.
\newblock \bibinfo{title}{Reviewer fatigue? why scholars decline to review
  their peers' work}.
\newblock \bibinfo{journal}{APSC} \bibinfo{volume}{48},
  \bibinfo{pages}{595--600}.
\newblock \DOIprefix\doi{10.1017/S1049096515000827}.
%Type = Article
\bibitem[{Budden et~al.(2008)Budden, Tregenza, Aarssen, Koricheva, Leimu and
  Lortie}]{BudTreAar08}
\bibinfo{author}{Budden, A.E.}, \bibinfo{author}{Tregenza, T.},
  \bibinfo{author}{Aarssen, L.W.}, \bibinfo{author}{Koricheva, J.},
  \bibinfo{author}{Leimu, R.}, \bibinfo{author}{Lortie, C.J.},
  \bibinfo{year}{2008}.
\newblock \bibinfo{title}{Double-blind review favours increased representation
  of female authors}.
\newblock \bibinfo{journal}{Trends in Ecology \& Evolution}
  \bibinfo{volume}{23}, \bibinfo{pages}{4--6}.
\newblock \DOIprefix\doi{10.1016/j.tree.2007.07.008}.
%Type = Article
\bibitem[{David~Pontille(2014)}]{Pontille2014}
\bibinfo{author}{David~Pontille, D.T.}, \bibinfo{year}{2014}.
\newblock \bibinfo{title}{The blind shall see! the question of anonymity in
  journal peer review}.
\newblock \bibinfo{journal}{Ada: A Journal of Gender, New Media, and
  Technology} \bibinfo{volume}{4}.
\newblock \DOIprefix\doi{10.7264/N3542KVW}.
%Type = Misc
\bibitem[{{Faculty of 1000 Ltd.}(2017)}]{F1000}
\bibinfo{author}{{Faculty of 1000 Ltd.}}, \bibinfo{year}{2017}.
\newblock \bibinfo{title}{F1000research}.
\newblock \bibinfo{howpublished}{\url{https://f1000research.com/about}}.
\newblock \bibinfo{note}{As of 2017-05-02}.
%Type = Article
\bibitem[{Ferreira et~al.(2016)Ferreira, Bastille-Rousseau, Bennett, Ellington,
  Terwissen, Austin, Borlestean, Boudreau, Chan, Forsythe, Hossie, Landolt,
  Longhi, Otis, Peers, Rae, Seguin, Watt, Wehtje and Murray}]{ferreira2016}
\bibinfo{author}{Ferreira, C.}, \bibinfo{author}{Bastille-Rousseau, G.},
  \bibinfo{author}{Bennett, A.}, \bibinfo{author}{Ellington, E.},
  \bibinfo{author}{Terwissen, C.}, \bibinfo{author}{Austin, C.},
  \bibinfo{author}{Borlestean, A.}, \bibinfo{author}{Boudreau, M.},
  \bibinfo{author}{Chan, K.}, \bibinfo{author}{Forsythe, A.},
  \bibinfo{author}{Hossie, T.}, \bibinfo{author}{Landolt, K.},
  \bibinfo{author}{Longhi, J.}, \bibinfo{author}{Otis, J.},
  \bibinfo{author}{Peers, M.}, \bibinfo{author}{Rae, J.},
  \bibinfo{author}{Seguin, J.}, \bibinfo{author}{Watt, C.},
  \bibinfo{author}{Wehtje, M.}, \bibinfo{author}{Murray, D.},
  \bibinfo{year}{2016}.
\newblock \bibinfo{title}{The evolution of peer review as a basis for
  scientific publication: directional selection towards a robust discipline}.
\newblock \bibinfo{journal}{Biol Rev Camb Philos Soc} \bibinfo{volume}{91},
  \bibinfo{pages}{597--610}.
\newblock \DOIprefix\doi{10.1111/brv.12185}.
%Type = Article
\bibitem[{Hettyey et~al.(2012)Hettyey, Griggio, Mann, Raveh, Schaedelin,
  Thonhauser, Thoss, van Dongen, White, Zala and Penn}]{Hettyey2012}
\bibinfo{author}{Hettyey, A.}, \bibinfo{author}{Griggio, M.},
  \bibinfo{author}{Mann, M.}, \bibinfo{author}{Raveh, S.},
  \bibinfo{author}{Schaedelin, F.}, \bibinfo{author}{Thonhauser, K.},
  \bibinfo{author}{Thoss, M.}, \bibinfo{author}{van Dongen, W.},
  \bibinfo{author}{White, J.}, \bibinfo{author}{Zala, S.},
  \bibinfo{author}{Penn, D.}, \bibinfo{year}{2012}.
\newblock \bibinfo{title}{Peerage of science: will it work?}
\newblock \bibinfo{journal}{Trends Ecol Evol} \bibinfo{volume}{27},
  \bibinfo{pages}{189–190}.
\newblock \DOIprefix\doi{10.1016/j.tree.2012.01.005}.
%Type = Article
\bibitem[{Jubb(2016)}]{jubb2016}
\bibinfo{author}{Jubb, M.}, \bibinfo{year}{2016}.
\newblock \bibinfo{title}{Peer review: The current landscape and future trends:
  Peer review landscape}.
\newblock \bibinfo{journal}{Learned Publishing} \bibinfo{volume}{29},
  \bibinfo{pages}{13--21}.
\newblock \DOIprefix\doi{10.1002/leap.1008}.
%Type = Article
\bibitem[{Laband and Piette(1994)}]{LabPie94}
\bibinfo{author}{Laband, D.N.}, \bibinfo{author}{Piette, M.J.},
  \bibinfo{year}{1994}.
\newblock \bibinfo{title}{A citation analysis of the impact of blinded peer
  review}.
\newblock \bibinfo{journal}{JAMA} \bibinfo{volume}{272},
  \bibinfo{pages}{147--149}.
\newblock \DOIprefix\doi{10.1001/jama.1994.03520020073020}.
%Type = Article
\bibitem[{Mulligan et~al.(2013)Mulligan, Hall and Raphael}]{mulligan2013}
\bibinfo{author}{Mulligan, A.}, \bibinfo{author}{Hall, L.},
  \bibinfo{author}{Raphael, E.}, \bibinfo{year}{2013}.
\newblock \bibinfo{title}{Peer review in a changing world: An international
  study measuring the attitudes of researchers}.
\newblock \bibinfo{journal}{Journal of the American Society for Information
  Science and Technology} \bibinfo{volume}{64}, \bibinfo{pages}{132--161}.
\newblock \URLprefix \url{http://dx.doi.org/10.1002/asi.22798},
  \DOIprefix\doi{10.1002/asi.22798}.
%Type = Misc
\bibitem[{Nature(2006)}]{Nature06}
\bibinfo{author}{Nature}, \bibinfo{year}{2006}.
\newblock \bibinfo{title}{Nature's peer review debate}.
\newblock
  \bibinfo{howpublished}{\url{http://www.nature.com/nature/peerreview/debate/index.html}}.
%Type = Article
\bibitem[{Prechelt et~al.(2017)Prechelt, Graziotin and Méndéz}]{Prechelt2017}
\bibinfo{author}{Prechelt, L.}, \bibinfo{author}{Graziotin, D.},
  \bibinfo{author}{Méndéz, D.}, \bibinfo{year}{2017}.
\newblock \bibinfo{title}{Open science repository for 'on the status and future
  of peer review in software engineering'}.
\newblock \bibinfo{journal}{figshare} \URLprefix
  \url{https://doi.org/10.6084/m9.figshare.5086357},
  \DOIprefix\doi{10.6084/m9.figshare.5086357}.
%Type = Misc
\bibitem[{Publons(2017)}]{Publons}
\bibinfo{author}{Publons}, \bibinfo{year}{2017}.
\newblock \bibinfo{title}{Recognising peer review. speeding up research.}
\newblock \bibinfo{howpublished}{\url{https://publons.com}}.
\newblock \bibinfo{note}{As of 2017-05-02}.
%Type = Article
\bibitem[{Ralph(2016)}]{ralph2016}
\bibinfo{author}{Ralph, P.}, \bibinfo{year}{2016}.
\newblock \bibinfo{title}{Practical suggestions for improving scholarly peer
  review quality and reducing cycle times}.
\newblock \bibinfo{journal}{Communications of the Association for Information
  Systems} \bibinfo{volume}{38}, \bibinfo{pages}{13}.
\newblock \URLprefix \url{http://aisel.aisnet.org/cais/vol38/iss1/13}.
%Type = Article
\bibitem[{Ross-Hellauer et~al.(2017)Ross-Hellauer, Deppe and
  Schmidt}]{RosDepSch17}
\bibinfo{author}{Ross-Hellauer, T.}, \bibinfo{author}{Deppe, A.},
  \bibinfo{author}{Schmidt, B.}, \bibinfo{year}{2017}.
\newblock \bibinfo{title}{http://doi.org/10.5281/zenodo.570864}.
\newblock \bibinfo{journal}{Zenodo} \URLprefix
  \url{http://doi.org/10.5281/zenodo.570864}. \bibinfo{note}{preprint}.
%Type = Incollection
\bibitem[{Schwarz and Bohner(2001)}]{Schwarz2001}
\bibinfo{author}{Schwarz, N.}, \bibinfo{author}{Bohner, G.},
  \bibinfo{year}{2001}.
\newblock \bibinfo{title}{The construction of attitudes}, in:
  \bibinfo{editor}{Tesser, A.}, \bibinfo{editor}{Schwarz, N.} (Eds.),
  \bibinfo{booktitle}{Blackwell Handbook of Social Psychology: Intraindividual
  Processes}. \bibinfo{publisher}{Blackwell Publishers Inc.},
  \bibinfo{address}{Malden, Massachusetts, USA}, pp. \bibinfo{pages}{436--457}.
\newblock \DOIprefix\doi{10.1002/9780470998519.ch20}.
%Type = Article
\bibitem[{ScienceOpen(2017)}]{ScienceOpen}
\bibinfo{author}{ScienceOpen, I.}, \bibinfo{year}{2017}.
\newblock \bibinfo{title}{Peer review guidelines}.
\newblock \bibinfo{journal}{ScienceOpen} \URLprefix
  \url{http://about.scienceopen.com/peer-review-guidelines/}.
%Type = Article
\bibitem[{Smith(2006)}]{Smith2006}
\bibinfo{author}{Smith, R.}, \bibinfo{year}{2006}.
\newblock \bibinfo{title}{Peer review: a flawed process at the heart of science
  and journals.}
\newblock \bibinfo{journal}{J R Soc Med} \bibinfo{volume}{99},
  \bibinfo{pages}{178–182}.
\newblock \DOIprefix\doi{10.1258/jrsm.99.4.178}.
%Type = Book
\bibitem[{Strauss and Corbin(1990)}]{StrCor90}
\bibinfo{author}{Strauss, A.L.}, \bibinfo{author}{Corbin, J.M.},
  \bibinfo{year}{1990}.
\newblock \bibinfo{title}{Basics of Qualitative Research: Grounded Theory
  Procedures and Techniques}.
\newblock \bibinfo{publisher}{{SAGE}}.
%Type = Article
\bibitem[{Warne(2016)}]{warne2016}
\bibinfo{author}{Warne, V.}, \bibinfo{year}{2016}.
\newblock \bibinfo{title}{Rewarding reviewers--sense or sensibility? a wiley
  study explained}.
\newblock \bibinfo{journal}{Learned Publishing} \bibinfo{volume}{29},
  \bibinfo{pages}{41--50}.
\newblock \DOIprefix\doi{10.1002/leap.1002}.
%Type = Book
\bibitem[{Weller(2001)}]{weller2001editorial}
\bibinfo{author}{Weller, A.C.}, \bibinfo{year}{2001}.
\newblock \bibinfo{title}{Editorial Peer Review: Its Strengths and Weaknesses}.
\newblock \bibinfo{publisher}{Information Today Inc}.
%Type = Article
\bibitem[{Zaharie and Osoian(2016)}]{zaharie2016}
\bibinfo{author}{Zaharie, M.A.}, \bibinfo{author}{Osoian, C.L.},
  \bibinfo{year}{2016}.
\newblock \bibinfo{title}{Peer review motivation frames: A qualitative
  approach}.
\newblock \bibinfo{journal}{European Management Journal} \bibinfo{volume}{34},
  \bibinfo{pages}{69–79}.
\newblock \DOIprefix\doi{10.1016/j.emj.2015.12.004}.

\end{thebibliography}

\end{document}